\documentclass[journal=acphoton,manuscript=article]{achemso}

\usepackage[version=3]{mhchem} 

\author{Alexander Neuhaus}
\email{alexander.neuhaus@uni-due.de}
\author{Pascal Dreher}
\altaffiliation{Present address: Institute of Physical and Theoretical Chemistry University of Würzburg, 97074 Würzburg, Germany.}
\author{Philipp Gessler}

\affiliation[1]{
 Faculty of Physics and Center for Nanointegration, Duisburg-Essen (CENIDE), University of Duisburg-Essen, 47048 Duisburg, Germany
}%

\author{Bettina Frank}
\affiliation[2]{
4th Physics Institute, Research Center SCoPE, and Integrated Quantum Science and Technology Center, University of Stuttgart, Stuttgart, Germany.
}%

\author{Timothy J. Davis}
\affiliation[1]{
 Faculty of Physics and Center for Nanointegration, Duisburg-Essen (CENIDE), University of Duisburg-Essen, 47048 Duisburg, Germany
}%
\alsoaffiliation[2]{
 4th Physics Institute, Research Center SCoPE, and Integrated Quantum Science and Technology Center, University of Stuttgart, Stuttgart, Germany.
}%
\alsoaffiliation[3]{
 School of Physics, University of Melbourne, Parkville, Victoria, Australia
}%
\author{Harald Giessen}
\affiliation[2]{
 4th Physics Institute, Research Center SCoPE, and Integrated Quantum Science and Technology Center, University of Stuttgart, Stuttgart, Germany.
}%

\author{Karin Everschor-Sitte}
\author{Frank-J. Meyer zu Heringdorf}
\email{meyerzh@uni-due.de}
\affiliation[1]{
 Faculty of Physics and Center for Nanointegration, Duisburg-Essen (CENIDE), University of Duisburg-Essen, 47048 Duisburg, Germany
}%

\title{Analysis of the Topology of a Plasmonic Target-Skyrmion Texture}


\begin{document}

\begin{tocentry}

\includegraphics[width=8.25cm]{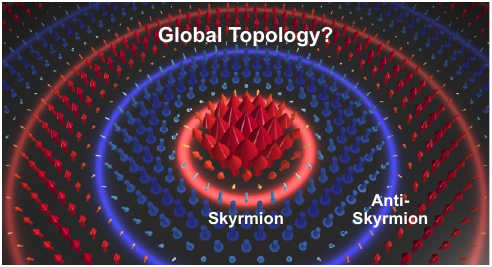}

\end{tocentry}

\begin{abstract}
Topological concepts are frequently used to describe structured optical fields, including plasmonic near fields. Topological descriptions in terms of skyrmion numbers implicitly assume the compactness of the underlying manifold.  Even when skyrmion-like textures appear locally, the compactness is usually not  fulfilled in extended optical fields. Here, we use photoemission electron microscopy to investigate a plasmonic nano-focus that exhibits a sequence of radially extending alternating skyrmion and antiskyrmion textures. The full spatio-temporal reconstruction of the electric field vectors and their topology is accessible by vector polarimetry. The experiments confirm the expected oscillatory behavior of the skyrmion number and demonstrate that a global skyrmion number cannot be assigned in such non-compact fields.  
\end{abstract}

\section{\label{sec:Intro}Introduction}
Structured optical near fields enable sub-wavelength control of light \cite{Frank2017}, and complex topological features have been identified within their respective electromagnetic field textures \cite{Dai2020, Ghosh2021, Du2019, Dai2022, Tsesses2018, Davis2020, Schwab2025, Dreher2024}. 
On noble metal surfaces, Surface Plasmon Polaritons (SPPs) provide control over electromagnetic fields on a nanoscale. Accordingly, 
topological structures such as skyrmions \cite{Tsesses2018, Du2019, Davis2020}, merons \cite{Dai2020,Ghosh2021, Dreher2024}, and phase vortices \cite{Tsesses2019, Bauer2023, Davis2023, Gorodetski2008, Kim2010} have been reported. Moreover, topology manifests itself not only in SPP electric fields \cite{Tsesses2018, Davis2020, Schwab2025}, but also in derived vector fields like the spin angular momentum  \cite{Du2019,Dai2020,Ghosh2021,Dai2022,Dreher2024}. 

It is important to emphasize that the application of well-defined topological concepts to local vectorial field textures is not at all unproblematic. This can be exemplified using a simple skyrmionic field texture. Skyrmions can be understood as vector fields mapped onto a sphere, with the skyrmion number counting how many times the sphere is covered by the field vectors. The concept requires that the vector field is defined on a compact spatial domain. Such compactness is naturally fulfilled if the field forms a periodic lattice, where the unit cell of the lattice is already compact in itself. In non-periodic field textures a construction known as Alexandroff compactification can be applied \cite{Alexandroff1924}, where a fixed vector direction at infinity is assumed. The well-known case of a magnetic skyrmion surrounded by a ferromagnetic background is an example for such compactification \cite{Koraltan2026}. In the case of SPPs, which represent a three-dimensional vector field defined on a two-dimensional plane, it would imply that, at infinity, all vectors must point in the same direction so that they map to a single point on the sphere. However, extended SPP near fields generally do not satisfy this requirement. Instead of approaching a fixed state at infinity, SPP fields remain oscillatory and can even contain amplitude zeroes where the vector direction becomes ill-defined across the observed region. Accordingly, the boundary of such field distribution cannot be treated as a single well-defined point of the mapping. Often, the analysis is then restricted to a sub-region of the field, describing a local skyrmionic texture \cite{Du2019, Dai2022} or a local meronic texture \cite{Dai2020, Dreher2024}. The  assignment of a global skyrmion number to the vector field is in such cases impossible.  

Here, we explore this limitation in assigning a global skyrmion number by investigating a plasmonic nano-focus with zero orbital angular momentum \cite{Frank2017, Dreher2022}; a vectorial field texture with cylindrical symmetry that is commonly referred to as a target skyrmion \cite{Tian2023}.  Using polarimetric photoemission electron microscopy \cite{Davis2020, Dreher2024} with deep sub-wavelength spatial and sub-cycle temporal resolution, we reconstruct the full vectorial electric field and directly access the skyrmion density in time. We show that the integration of the skyrmion density as a function of the radius does not converge to a stable skyrmion number,  but that the skyrmion number oscillates as a function of the integration radius. Our results confirm that the presence of a locally skyrmion-like texture does not imply a globally protected topological invariant.

\section{\label{sec:Results} Experimental Setup}
Access to the time-resolved electric field vectors is essential for a topological analysis of a plasmonic nano-focus, because the related skyrmion density depends on the instantaneous vector configuration of the field. We therefore need to measure the full vectorial electric field of the plasmonic nano-focus as a first step of the topological analysis.

\begin{figure*}
    \centering
    \includegraphics[width=\textwidth]{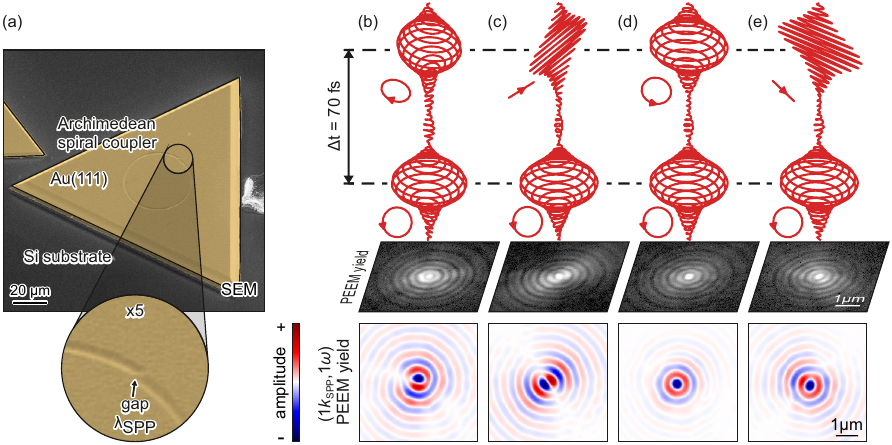}
    \caption{Experimental setup and measurement scheme. (a) scanning electron micrograph of the investigated Au(111) platelet with an Archimedean spiral milled into the surface. (b)-(e) reconstructed pulse trains (electric field trajectory drawn as solid line) with fixed left circular pump polarization and varying probe polarization at a delay of $70\,\mathrm{fs}$. The pulse trains impinge on the sample to create the raw PEEM images shown in logarithmic grayscale in a perspective view. The pulse trains have been reconstructed from a second harmonic dispersion scan \cite{Miranda2012,Geib2019}. The spatio-temporally Fourier filtered PEEM images are shown at the bottom of panels (b)-(e), which represent the projection of the SPP onto the different probe polarizations.}
    \label{fig:1}
\end{figure*}

Polarimetric photoemission electron microscopy (PEEM), introduced by Dreher et al. \cite{Dreher2024}, provides us with both the deep sub-wavelength spatial resolution required to resolve the nano-focus and the sub-cycle temporal resolution necessary to capture the temporal oscillation of the plasmon field. Further experimental details are provided in the Methods section. In brief, SPPs are optically excited by a femtosecond laser pump pulse at a grating coupler that provides the required momentum matching. To generate the nano-focus, an Archimedean spiral coupler with an opening of one SPP wavelength \cite{Kim2010} is milled into a Au(111) platelet \cite{Radha2010, Radha2011} (Fig.~\ref{fig:1}a). The sample is illuminated with a left circularly polarized pump pulse at a central wavelength of $800\,\mathrm{nm}$. Concentric SPP wave-fronts are created at the spiral coupler \cite{Dreher2022} through spin-angular-momentum to orbital-angular-momentum conversion\cite{Spektor2017, Spektor2019}.  The nano-focus is formed about $75\,\mathrm{fs}$ after the excitation. The SPP wavelength of $783\,\mathrm{nm}$ is determined by the Au(111) substrate, the surrounding vacuum, and the laser frequency.

A second, time-delayed and phase-stable probe pulse with variable polarization \cite{Neuhaus2025} is directed onto the sample to access the electric field configuration of the SPPs. In combination with the pump pulse it forms the phase-locked pulse trains shown at the top of Fig.~\ref{fig:1}b-e. The interference between the SPP field and the probe pulse drives a two-photon photoemission process \cite{Kubo2007,Dreher2024a}, and the emitted electrons are subsequently imaged in a low-energy electron microscope with a time-integrating detector \cite{Janoschka2021}. By systematically varying the polarization of the probe pulse, multiple projections of the SPP electric field are recorded, enabling the reconstruction of the full vectorial field above the surface. 
The polarization-resolved raw PEEM images of the plasmonic nano-focus are shown in the middle of each of the panels (b)-(e) of Fig.~\ref{fig:1}, revealing a pronounced dependence on the probe polarization. Repeating these measurements as a function of the pump-probe delay allows the reconstruction of the full temporal evolution of the SPP field.

\begin{figure*}
    \centering
    \includegraphics[width=\textwidth]{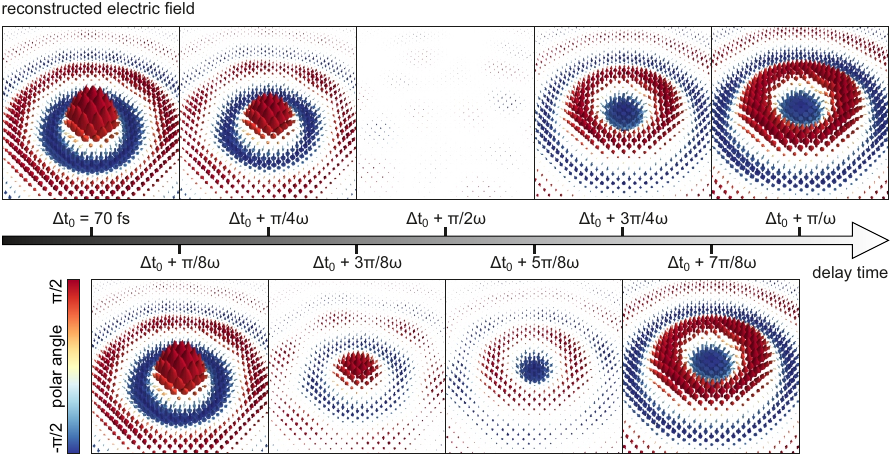}
    \caption{Reconstructed electric field vectors as a function of delay time for half an optical cycle ($T=2.66\, \mathrm{fs}$). The central skyrmion-like texture switches to an antiskyrmion-like texture after the zero crossing.}
    \label{fig:2}
\end{figure*}

In the two-photon photoemission process, the measured electron yield contains additional contributions from plasmoemission \cite{Podbiel2017} and pure photoemission. These unwanted  signals are removed by spatio-temporal Fourier filtering \cite{Spektor2019, Davis2020}, isolating the first harmonic of the SPP field and thereby extracting the projection of the electric field onto the probe polarization, as shown in the bottom images in Figs.~\ref{fig:1}b-e. Since the probe pulse is incident normal to the surface \cite{Kahl2018}, only the in-plane component of the SPP field is directly accessible in the experiment \cite{Podbiel2016}. The full vectorial electric field is reconstructed by combining measurements for different probe polarizations using an iterative algorithm that accounts for nonlinearities in the photoemission process within the nano-focus \cite{Dreher2024}. Within this reconstruction, the out-of-plane component is retrieved by enforcing the divergence-free nature of the electric field in combination with its evanescent decay away from the surface \cite{Davis2020, Dreher2024}. 

\section{Results and Discussion}
The reconstructed electric field of the nanofocus close to the temporal overlap is shown in Fig.~\ref{fig:2}. As expected, the nanofocus exhibits a cylindrical symmetry. The different panels in Fig.~\ref{fig:2} show the electric field vectors at different delay times within half an optical cycle, beginning at a pump–probe delay of $\Delta t_0=70\,\mathrm{fs}$ after excitation. The experimental data at $\Delta t_0=70\,\mathrm{fs}$ reveals a pronounced out-of-plane field component at the center of the nano-focus, accompanied by a radially oriented in-plane component, locally forming a skyrmion-like texture. The entire electric field oscillates at the central frequency $\omega$ of the SPP and all vectors reverse their orientation every half optical cycle, passing through zero at every sign reversal (cf. the panel at $\Delta t = \Delta t_0 + \pi/2\omega$ in the center of Fig.~\ref{fig:2}).

\begin{figure*}
    \centering
    \includegraphics[width=130.7mm]{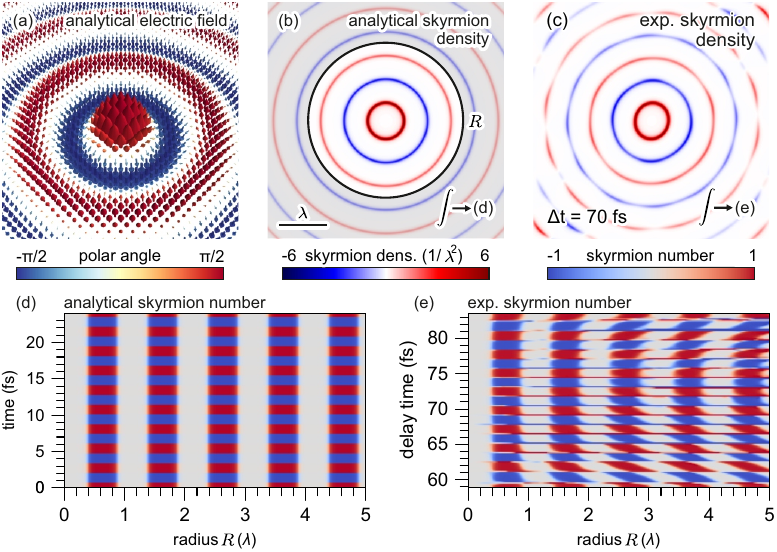}
    \caption{Topology of the plasmonic nano-focus. (a) analytical electric field distribution with a central skyrmion-like texture. (b) analytical skyrmion density $q$ with concentric rings of alternating sign. (c) skyrmion density $q$ extracted from the experimental data in Fig.~\ref{fig:2}. By integrating the skyrmion density $q$ within the circular mask of radius $R$, the skyrmion number $S$ plotted in panel (d) and (e) is obtained from the analytical model in (b) and the experimental data in (c), respectively.}
    \label{fig:3}
\end{figure*}

While the reconstructed electric field already reveals the skyrmion-like texture, its topology and the behavior of the skyrmion number are most easily understood within an analytical model. The experimental field configuration in Fig.~\ref{fig:2} can be approximated in a continuous wave limit by a cylindrical harmonic mode \cite{Davis2017}. The cylindrical symmetry allows for a description of field vectors in cylindrical coordinates with radius $r$, height $z$, and angle $\varphi$. For a wave at frequency $\omega$ with wavenumber $k$, the vectorial electric field directly above the surface ($z=0^+$) is given by
\begin{equation}
    \vec{E}(t,r,\varphi,z=0^+)=E_0 \cos{\omega t} \,
    \left(
    J_0(kr)\hat{e}_z+\eta \,J_1(kr)\hat{e}_r(\varphi)
    \right)
    \label{eq:electric_field}
\end{equation}
where $J_{\mathrm{n}}$ is the Bessel function of the first kind of order $\mathrm{n}$. The unit vectors in cylindrical coordinates are denoted as $\hat{e}_i$. The ratio $\eta$ describes the relative strength of the in-plane and out-of-plane components of the electric field that follows from the dielectric constants. Variation of $\eta$  continuously changes the texture from predominantly out-of-plane (bubble-like $\eta<1$) to a continuously rotating (Néel-like $\eta>1$) texture \cite{Schwab2026}. For the SPPs in the experiment, the out-of-plane component exceeds the in-plane component above the surface by a factor of five ($\eta\approx1/5$). 
A vector representation of this analytically calculated electric field is shown in Fig.~\ref{fig:3}a at a time comparable to $\Delta t_0$ in the experiment in Fig.~\ref{fig:2}.
The shown field configuration has been interpreted as a target skyrmion\cite{Zheng2017}, i.e., an alternating sequence of skyrmions and antiskyrmions \cite{Tian2023}. 


This alternating sequence can be easily understood by considering the skyrmion density $q$ of the field. It measures the solid angle density spanned by the normalized field vectors $\hat{n}$ and is given by

\begin{equation}
    q(r,\varphi) =
    \frac{1}{4\pi r}\,
    \hat{n}(r,\varphi)\cdot
    \left(
    \partial_r \hat{n}(r,\varphi)
    \times
    \partial_\varphi \hat{n}(r,\varphi)
    \right).
    \label{eq:skyrm_dens_operator}
\end{equation}

The skyrmion density from the analytically computed electric field vectors for a fixed time is shown in Fig.~\ref{fig:3}b, and the related skyrmion density from the experimental data is shown in Fig.~\ref{fig:3}c. In both cases, theory and experiment, the skyrmion density is localized on concentric rings with alternating sign, revealing the bubble-like character of the local skyrmionic texture ($\eta \approx 1/5$). At the time of maximum out-of-plane field in the center, the innermost ring carries positive topological charge, while the second ring exhibits the same charge with opposite sign, fully compensating the first. This alternating pattern extends over multiple wavelengths without any indication of convergence within the experimentally accessible field of view. We would like to emphasize that the calculation of the skyrmion density from experimental data involves several spatial derivatives of the reconstructed electric field, and that the polarimetric PEEM method is able to clearly resolve it.

The skyrmion number of the vector field follows from integrating the skyrmion density (cf. Fig.~\ref{fig:3}b,c) over a selected area. Considering the cylindrical symmetry of the field, this area is naturally chosen within a circular boundary of radius $R$. The alternating sequence of positively and negatively charged rings naturally make the skyrmion number $R$-dependent. This raises the question of whether a global well-defined skyrmion number can be assigned to the field at all. 
Applying the general requirements for a well-defined global skyrmion number, integer-valued skyrmion numbers on a two-dimensional plane require the field to approach a uniform background at infinity. Only under this condition can the mapping be compactified, allowing the skyrmion number to serve as a global invariant within the homotopy group $\pi_2 (S^2)$. For the field considered here, however, such a compactification is not possible: although the Bessel functions decay as $1/\sqrt{r}$, this decay is removed upon the necessary normalization of field vectors in the calculation of the skyrmion density, leaving a purely oscillatory behavior. Accordingly, the analytically derived skyrmion density in Fig.~\ref{fig:3}b shows concentric rings with alternating sign. As a result, the field does not converge to a uniform background at large distances, fundamentally violating the concept of a well-defined global skyrmion number.

The problem of compactification is further illustrated by plotting the skyrmion number as a function of time and integration radius $R$ in Fig.~\ref{fig:3}d. After the integration of Eq. \ref{eq:skyrm_dens_operator} (see Supplementary Material for derivation), one obtains the skyrmion number 

\begin{equation}
   S(R)= \mathrm{sgn}(\cos{\omega t})\frac{1}{2} \,\left(1-\frac{J_0 (kR)}{\sqrt{J_0^2 (kR)+\eta^2 J_1^2 (kR)}} \right). 
   \label{eq:Skyrmionnumber}
\end{equation}

Rather than converging to a constant value, Eq.~\ref{eq:Skyrmionnumber} shows that the skyrmion number remains strongly oscillatory with increasing radius and in time. Consequently, the assigned topological charge depends explicitly on the choice of integration boundary and time.  The skyrmion number thus does not constitute a topological invariant of the field. 

The spatio-temporal dependence of the skyrmion number in the pulsed-illumination experiment (Fig.~\ref{fig:3}e) agrees well with the analytical prediction of the continuous-wave model shown in Fig.~\ref{fig:3}d. In the experiment, the SPP pulses temporally overlap at the nanofocus at $\Delta t_0 = 75\,\mathrm{fs}$. Around this delay, the experimental pattern is closest to the stationary continuous-wave limit. At  delay times preceding the temporal overlap, the SPP pulses are still propagating towards the center of the structure, leading to a shear of the skyrmion-number domains at larger radii as a function of delay time. The same is true for the situation after temporal overlap, where the SPP pulses propagate away from the nanofocus and the shear pattern is reversed.

\begin{figure}
    \centering
    \includegraphics[width=0.5\linewidth]{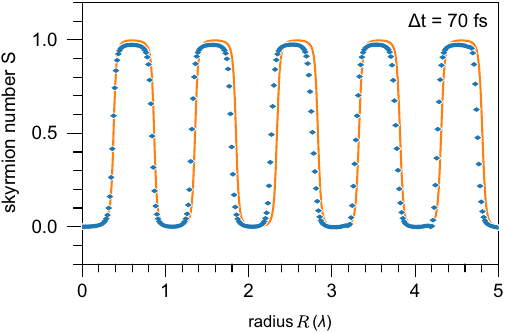}
    \caption{Comparison of experimentally determined (blue dots) and analytically predicted (orange line) skyrmion number $S$ as a function of radius at a delay time of $\Delta t=70\,\mathrm{fs}$.}
    \label{fig:4}
\end{figure}

To examine the radial dependence of the skyrmion number in more detail, Fig.~\ref{fig:4} shows a line cut through the analytical and experimental skyrmion-number maps in Figs.~\ref{fig:3}d,e at the experimental delay time of $\Delta t = 70\,\mathrm{fs}$. The experimental data closely follows the continuous-wave prediction and exhibits a sequence of nearly integer-valued plateaus near $S=1$ and $S=0$. Small deviations from the ideal integer values arise primarily from barrel distortions within the electron optics of the microscope, which slightly shift the radii at which the plateaus occur. Nevertheless, the close agreement between experiment and calculation confirms the existence of locally quantized values of the skyrmion number within the oscillatory texture.

The origin of these nearly integer-valued plateaus can be understood from the behavior of the field along the circular integration boundary. These special radii correspond to the zeros of the first-order Bessel function $J_1(kR)$. At these radii, the field on the circular integration boundary becomes purely out-of-plane, such that all boundary vectors point in the same direction. The boundary therefore maps to a single point on the unit sphere, allowing the integration domain to be compactified and yielding a locally well-defined integer-valued skyrmion number $S$.

As the electric field evolves in time, the field vectors reverse their orientation every half optical cycle. Due to the normalization involved in the calculation of the skyrmion density, this reversal translates into a sudden inversion of the topological texture, where the central skyrmion-like texture transforms into an antiskyrmion-like configuration at the zero crossing, as previously reported \cite{Davis2020,Tian2023, Schwab2025}. This can clearly be observed in the analytical prediction for the skyrmion number in Fig.~\ref{fig:3}d and the skyrmion number obtained from the polarimetric PEEM experiment in Fig.~\ref{fig:3}e. This temporal oscillation implies that even a locally defined topological charge is not conserved but instead periodically flips sign, averaging it to zero over a full optical cycle. We conclude that any assignment of a global topological invariant for this system remains questionable, since the necessary conditions of compactness and temporal stability are not fulfilled.

Even when considering averaged quantities at a fixed time, such as the Cesàro mean in the limit $R\to\infty$, the skyrmion number approaches a non-integer value of $S=1/2$. This value formally resembles the charge commonly associated with a meron-like texture \cite{Ezawa2011,Ghosh2021}. However, in the present case it arises only as an averaged quantity of the oscillatory skyrmion number and not as a globally defined topological charge. The skyrmion number itself remains boundary-dependent. Together with the persistent spatial oscillations of the field, this again demonstrates that no globally well-defined skyrmion number can be assigned to the plasmonic nano-focus. 

\section{\label{seq:conclusions}Conclusions}
The experimentally reconstructed field configuration of a plasmonic nano-focus exhibits a pronounced skyrmion-like vectorial texture. However, a detailed analysis shows that the corresponding skyrmion number depends explicitly on the choice of integration boundary and the time of observation. In the absence of a compact manifold, the conditions required for a well-defined topological invariant are not fulfilled, and the skyrmion number fails to provide a global classification of the field.

Several possible ways of addressing this limitation are conceivable, but each comes with substantial restrictions. One may restrict the integration to an inner domain in which the texture appears skyrmionic \cite{Tian2023}. This, however, introduces an arbitrary boundary and therefore does not provide a stringent global topological classification. A seemingly natural alternative would be to use the outer edge of the experimentally accessible region, which could be set by the grating coupler used to launch the SPPs. Yet this boundary is not topologically distinguished either, since the grating coupler also launches outward-propagating SPP components, so that the inner field would not describe the full topology.

Another possibility would be to consider confined plasmonic resonators. For the Archi\-medean spiral studied here, this route is not directly applicable, since the structure does not support the target skyrmion as a resonant eigenmode. Moreover, spatial confinement alone would not remove the more general difficulty that the instantaneous skyrmion number of the SPP field oscillates in time. The problem is therefore not merely experimental or geometrical, but reflects a more fundamental limitation of applying static real-space skyrmion invariants to oscillating optical fields.

Our results show that skyrmion-like field textures in plasmonic and optical systems must be interpreted with care. The presence of a locally nontrivial vectorial texture does not by itself imply a globally protected topological invariant. In particular, the target-skyrmion texture studied here is best understood as a skyrmion-like real-space configuration whose apparent skyrmion number is boundary-dependent, rather than as a globally protected skyrmion.

A different route is to seek invariants that are formulated for the global structure of wave fields rather than for a spatially truncated real-space domain. In a recent preprint \cite{Neuhaus2026}, we introduced a momentum-space linking number as a candidate for a global topological invariant for infinite vector wave fields. Applied to the target-skyrmion field considered here, this linking number is zero at all times and is independent of an arbitrarily chosen integration boundary.  The momentum-space linking number appears as a well-defined topological quantity, but it measures a different aspect of the field \cite{Neuhaus2026} than the skyrmion number does. The real-space skyrmion number, however, is not a well-defined global topological invariant for the plasmonic target skyrmion.

\section{\label{seq:methods}Methods}
Experiments were performed on a thermolytically grown Au(111) platelet \cite{Radha2010, Radha2011}. An Archi\-medean spiral coupler with a radius of $20\,\mathrm{\mu m}$ and an opening corresponding to one SPP wavelength was fabricated into the surface using focused $\mathrm{Au}^{2+}$-ion beam milling (Raith Ionline Plus). The sample was transferred into the ultra-high vacuum system of an ELMITEC SPELEEM III \cite{Schmidt1998} microscope. The Au(111) surface was prepared by repeated sputter-anneal cycles, yielding a clean and well-ordered surface \cite{Dreher2023}.
Time-resolved measurements were carried out using a mode-locked Ti:Sapphire laser delivering $<15\,\mathrm{fs}$ pulses at a repetition rate of $80 \,\mathrm{MHz}$ and a central wavelength of $800\,\mathrm{nm}$. Phase-stable pump-probe pulse pairs with controllable polarization were generated using an actively stabilized Mach-Zehnder interferometer \cite{Neuhaus2025}. The relative phase of the probe pulse and the temporal overlap were determined using a combination of spectral interference and polarimetry \cite{Neuhaus2025}. In the polarimetric PEEM experiment, SPPs were excited by the pump pulse and probed by a delayed pulse in a two-photon photoemission process.
To ensure a two-photon photoemission regime, the work function of the Au(111) surface was reduced by deposition of a sub-monolayer amount of Cs \cite{Petek2001}. The spatially resolved photoemission yield was recorded as a function of pump–probe delay and probe polarization, resulting in a four-dimensional dataset. The electron yield was processed by spatio-temporal Fourier filtering to isolate the first harmonic of the SPP, providing the projection of the electric field onto the probe polarization\cite{Davis2020}. The full vectorial electric field was reconstructed using an iterative algorithm that accounts for nonlinear photoemission contributions \cite{Dreher2024}. Exploiting the cylindrical symmetry of the system, residual artifacts due to non-uniform illumination were removed by symmetrization in the Fourier domain.

\begin{acknowledgement}

We acknowledge support from the ERC (Complexplas, 3DPrintedoptics) (B.F. and H.G.), DFG (grant no. SPP1391 Ultrafast Nanooptics (A.N., P.D., B.F., H.G., and F.-J.M.z.H.), CRC 1242 Non-Equilibrium Dynamics of Condensed Matter in the Time Domain (project no. 278162697-SFB 1242 (project B06 and B10) to A.N., P.G., P.D., K.E.-S. and F.-J.M.z.H.)), BMBF (Printoptics) (B.F. and H.G.), BW Stiftung (Spitzenforschung, Opterial) (B.F. and H.G.) and Carl-Zeiss Stiftung (B.F. and H.G.). T.J.D. acknowledges support from the MPI Guest Professorship Program and from the DFG (grant no. GRK2642 Photonic Quantum Engineers) for a Mercator Fellowship.

\end{acknowledgement}

\begin{suppinfo}

Supporting Information S1: Analytical derivation of the skyrmion number as a function of integration radius.

\end{suppinfo}

\bibliography{Refs_cleaned}

\clearpage
\appendix
\setcounter{figure}{0}
\renewcommand{\thefigure}{S\arabic{figure}}
\setcounter{equation}{0}
\renewcommand{\theequation}{S\arabic{equation}}

\section{Supporting Information: S1 Derivation of the Skyrmion Number}

This supplementary note will derive the skyrmion number as a function of the integration radius as discussed in the main manuscript. We start by considering the electric field  of the plasmonic nanofocus at $t=0$ in cylindrical coordinates given by Eq. 1 of the main manuscript. For the calculation of the Skyrmion number, we normalize electric field vectors
\begin{equation}
    \hat{n}(r,\varphi)=\frac{1}{\sqrt{(J_0^2(kr) + \eta^2 J_1^2(kr)}}
    \left(
    J_0(kr)\hat{e}_z+\eta \,J_1(kr)\hat{e}_r(\varphi)
    \right).
    \label{eq:norm_field}
\end{equation}

Utilizing the cylindrical symmetry of the vector field, it is useful to introduce the fields polar angle $\theta(r)$ as 

\begin{equation}
    \tan\theta(r)=\frac{\eta\, J_1(kr)}{J_0(kr)}.
    \label{eq:angle_definition}
\end{equation}

Using this angle, one can rewrite Eq.~\ref{eq:norm_field} to a much simpler form

\begin{equation}
    \hat{n}(r,\varphi)=\cos{\theta(r)}\,\hat{e}_z+\sin{\theta(r)}\,\hat{e}_r(\varphi).
    \label{eq:angle_field}
\end{equation}

The Skyrmion number is defined as the integral over the skyrmion density, which captures the local solid angle spanned by the normalized field vectors per area. In cylindrical coordinates, the skyrmion density $q$ is given by Eq.~2 of the main text.
Inserting the angle representation Eq.~\ref{eq:angle_field} into Eq.~2 of the main text yields the skyrmion density of the plasmonic nano-focus

\begin{equation}
\begin{split}
q(r,\varphi) 
&=\frac{1}{4\pi r}\,\left(\partial_r \, \theta(r)\right)\,\sin{\theta(r)}\\
&=-\frac{1}{4\pi r}\,\left(\partial_r\cos{\theta(r)}\right).
\label{eq:skyrm_dens}
\end{split}
\end{equation}

The skyrmion number can then be obtained by spatial integration of the skyrmion density. Following the cylindrical symmetry, the skyrmion density is integrated over a circular area with radius $R$ yielding the radius dependent skyrmion number 

\begin{equation}
\begin{split}
S(R) 
&=-\int_0^{2\pi}\mathrm{d}\varphi\int_0^R r  \frac{1}{4\pi r}\,\left(\partial_r \,\cos{\theta(r)}\right)\mathrm{d}r\\
&=\frac{1}{2}\left(\cos{(0)}-\cos{\theta(R)}\right).
\label{eq:skyrm_num}
\end{split}
\end{equation}

As $\theta(r)$ is given by Eq.~\ref{eq:angle_definition}, and both $J_0$ and $J_1$ oscillate as functions of $r$, the skyrmion winding number mirrors this oscillation and does not converge to a constant. Inserting Eq.~\ref{eq:angle_definition} explicitly into Eq.~\ref{eq:skyrm_num} yields Eq.~3 of the main manuscript

\begin{equation}
   S(R)=\frac{1}{2} \,\left(1-\frac{J_0 (kR)}{\sqrt{J_0^2 (kR)+\eta^2 J_1^2 (kR)}} \right). 
   \label{eq:Skyrmionnumber}
\end{equation}

This function remains highly oscillatory and does not converge for $R\to\infty$. Only a renormalization of the integral exists for $R\to\infty$, such as the Cesàro mean, as discussed in the main manuscript.

\end{document}